\documentclass{nwkltrs}

\usepackage{makecell,url,arydshln}

\sidecaptionvpos{figure}{c}

\begin{document}

\title{Parliamentary mention networks reflect the organisation and dynamics of Finnish political discourse}

\headertitle{Parliamentary mention networks}

\author[1]{Henna Poikkim\"aki}

\author[2,1]{Petri Leskinen}

\author[1,3]{Petter Holme}

\affil[1]{Department of Computer Science, Aalto University, 02150 Espoo, Finland}

\affil[2]{Department of Computer Science, University of Helsinki, 00100 Helsinki, Finland}

\affil[3]{Center for Computational Social Science, Kobe University, Kobe 657-8501, Japan}

\twocolumn[%
\begin{@twocolumnfalse}
  \vspace{-15mm}
\maketitle
\abstract{The primary purpose of the parliamentary discussion is to present and debate legislative drafts. Mentions of other Members of the Parliament (MPs) and interruptions of them represent the structure of the discourse. Name mentions can be automatically recognised in speech transcripts and linked to the relevant MPs. We construct networks of MPs based on name mentions in 96753 speeches delivered during the electoral terms 2015--2018 and 2019--2022 in the Finnish Parliament to study the structures of interaction within parliamentary discussion. The results show that the networks are not polarized, although there is slight bias towards mentioning one's own party members. Activity on the parliamentary floor, political seniority and minister status have a positive correlation with higher centrality scores of MPs. Our analysis is generalizable to different political arenas and is an interesting basis for comparison across regional assemblies or countries.}
\vspace{5mm}
\end{@twocolumnfalse}]
\thispagestyle{empty}

\section{Introduction}\label{sec_intro}

The primary important task of the Parliament of Finland---or, indeed, the parliament of any Western democracy---is enacting legislation. During plenary sessions, Parliament debates government proposals and sometimes Citizens' initiatives, turns them into legislation, and votes on approval. On Question Time, ministers have to answer questions from Members of Parliament (MPs). Although political debate speeches given during plenary sessions are sometimes considered ``political theater'' by the general public~\cite{pekonen_puhe_eduskunnassa_2011}, transcripts of legislative and parliamentary speeches have increasingly attracted the attention of researchers from different fields of study~\cite{abercrombie2020sentiment}. 

Many studies discuss the benefits of the activity on the parliamentary floor for politicians and political parties. For example, in the case of Belgian Parliaments, party loyalty and parliamentary activity, including activity on the parliamentary floor, can positively affect the expectations for re-election and promotion to higher parliamentary office~\cite{schobess-2022-parl-career}. In addition, visibility in traditional and social media can increase with parliamentary activity. There is a positive association between the number of legislative speeches and the media appearances of MPs in the United Kingdom and Norway, where especially senior MPs and MPs of the government party benefit from active debate on the parliamentary floor~\cite{yildirim-2023-parl-activities-media}. Although social media is a crucial tool for political parties to highlight issues and set the political agenda, the issues raised on social media often mirror those discussed in parliamentary settings~\cite{poljak-2025-speechtofeed}. On the other hand, social media like X or Bluesky can provide an instant and more individualised channel of communication for MPs, without being tied down by party politics~\cite{castanho-2022-twitter-vs-parliament}.

\begin{SCfigure*}[0.5]
    \centering
    \includegraphics[width=1.5\linewidth]{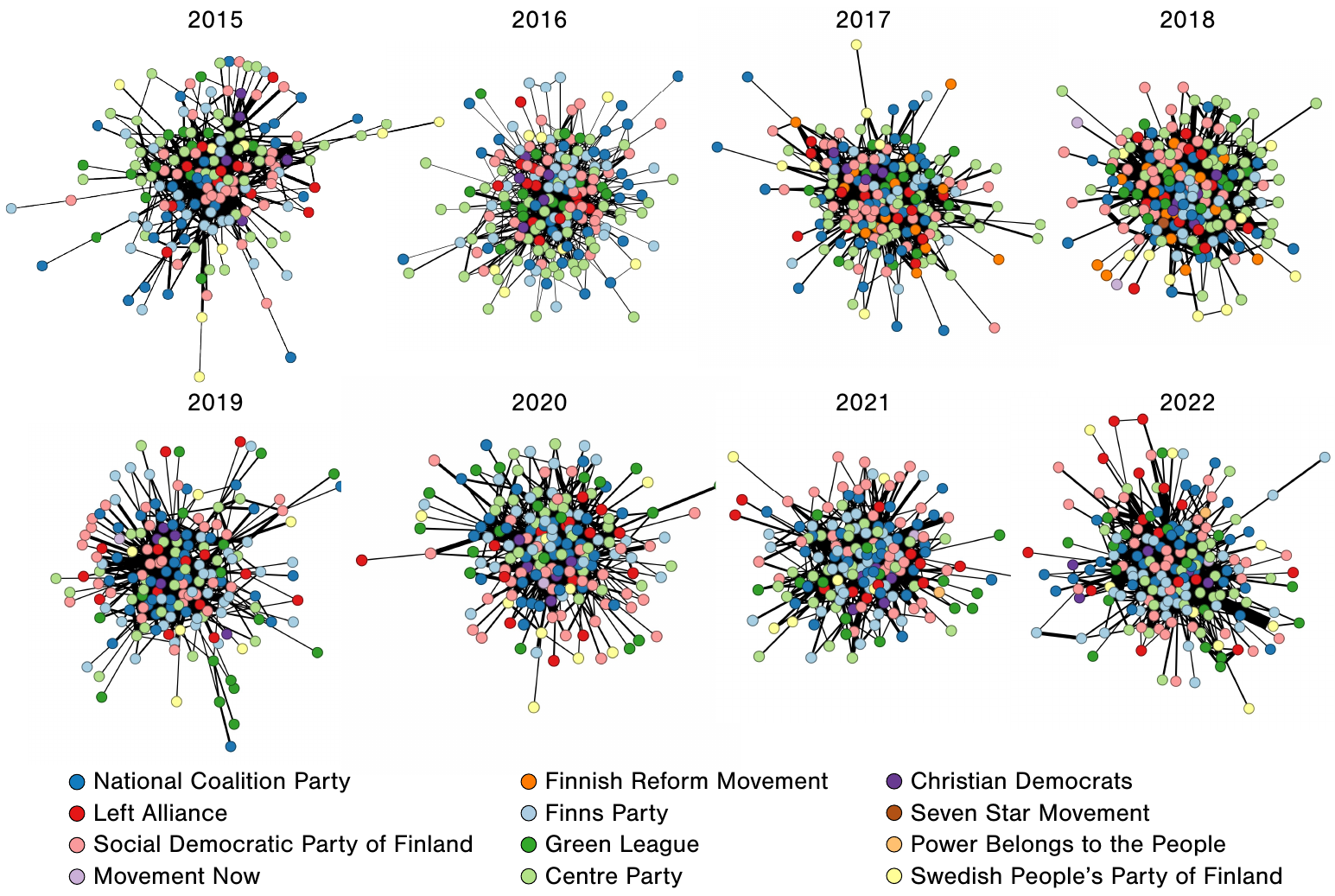}
    \caption{Networks for each parliamentary session after applying the disparity filter. The nodes are colored by parties. The network layout was done with the Gephi's Yifan Hu implementation.}
    \label{fig:networks}
\end{SCfigure*}

In Finland, every four years, 200 MPs are elected to the government, and the chairperson of the party with the most elected MPs typically becomes prime minister and begins negotiations for the government coalition. Coalition parties have to stay unified to govern together, but also have their own voice to attract voters~\cite{sagarzazu2017coalition}, while opposition parties challenge the government by pointing out the weaknesses of government policies and offering alternatives~\cite{demirkaya-2019-opposition}.
Support or opposition to other MPs and the groups they represent can be shown through interruptions to speeches and direct mentions of names. The mentions of other MPs and interruptions indicate potential debate~\cite{pekonen_puhe_eduskunnassa_2011}.

In this work, we aim to study the potential debate through name-mention networks, in which nodes are elected MPs connected by directed, weighted edges representing name mentions in parliamentary speeches~\cite{poikkimaki2022analyses}. We construct networks for each yearly parliamentary session during two contrasting electoral terms: 2015--2018 with centre-right government coalition and 2019--2022 with centre-left coalition. We utilise several network analysis methods to study the macro- and micro-level conversational structures of the networks and to compare results between parliamentary sessions and electoral terms. The ParliamentSampo~\cite{hyvonen_ps_data_2024} contains all speeches and speakers of the Parliament of Finland, along with relevant metadata, harmonised as linked open data, since the Parliament's founding in 1907. At the moment, since the 2015 parliamentary session, named entities, including mentions of people, have been recognised in speeches and linked to corresponding resources.

\begin{SCfigure*}[0.5]
    \centering
    \includegraphics[width=1.5\linewidth]{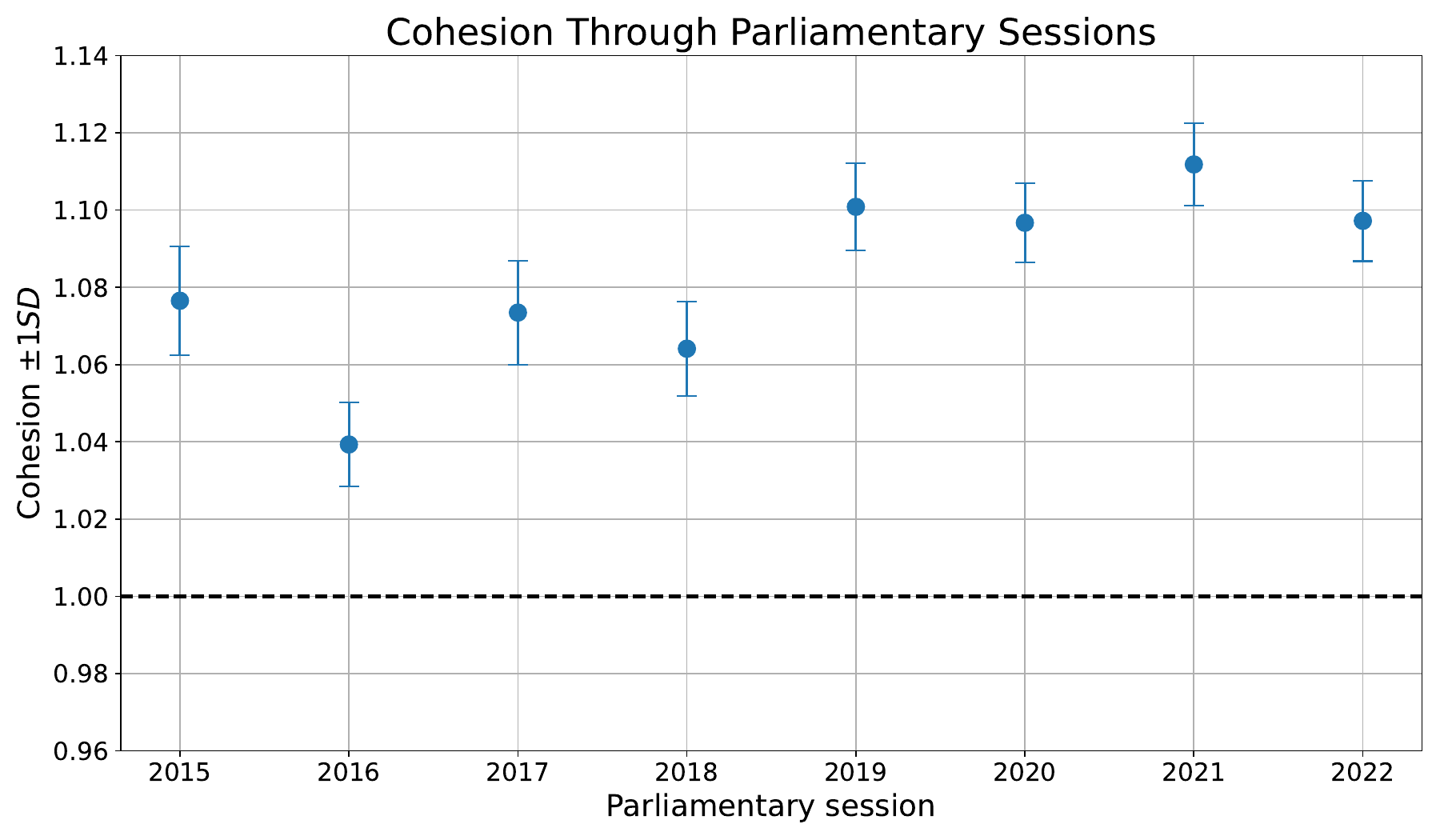}
    \caption{Cohesion for parliamentary sessions after the networks are simplified by the disparity filter. The p-value of 2016 is  $\approx 0.26$, and that of 2018 is $\approx0.015$, other p-values are $< 0.01$ ($N=1000$). Error bars represent one standard deviation.}
    \label{fig:coefficients}
\end{SCfigure*}

\begin{SCfigure*}[0.5]
    \centering
    \includegraphics[width=1.5\linewidth]{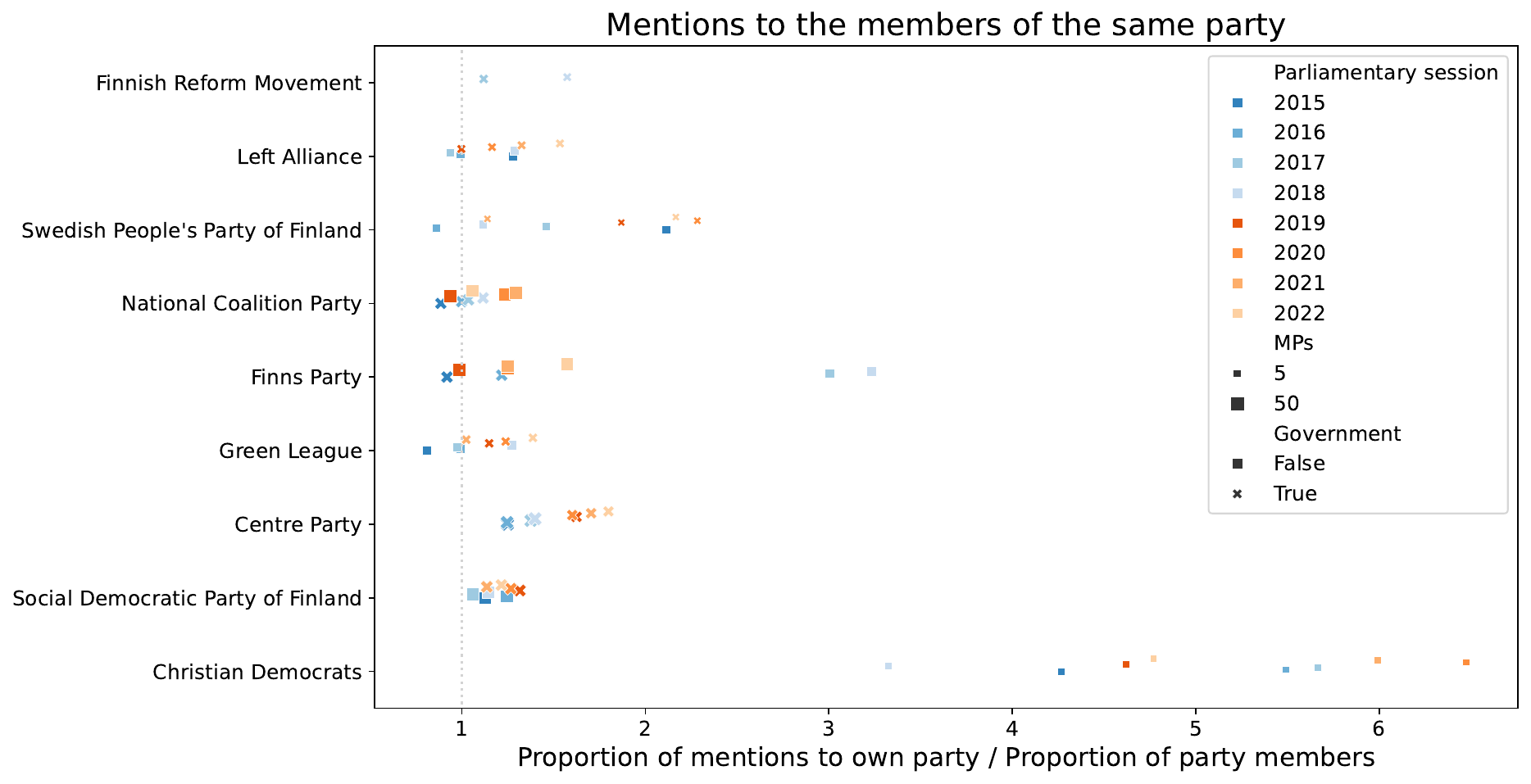}
    \caption{The proportion of mentions by party members directed towards the members of their own party divided by the proportion of the party members in the Parliament.}
    \label{fig:party_self_mentions}
\end{SCfigure*}

\section{Related work}

A wide range of computational methods have been adapted and developed for studying transcripts of parliamentary debates, ranging from NLP techniques such as sentiment analysis and topic detection to network analysis~\cite{abercrombie2020sentiment}. For example, networks of New Zealand Parliamentarians based on shared topics in parliamentary speeches and popularity of topics over time have been studied in~\cite{curran2018look}. Members of parties in the UK House of Commons have been classified according to their speeches, where the accuracy of the classification against party labels expresses the level of polarization over the years~\cite{peterson-2018-polarization}. Televised political debate has also been used to construct networks of politicians by recognising different events, such as adversarial or supportive interruptions, and further reconstructing them into relationships such as alliances, attacks, or ignorance~\cite{fuhse2023analyzing}.

Networks of parliamentarians based on name mentions in speeches given in the European Parliament have been studied in~\cite{walter_ep_mentions_2023}. Applying text analysis and dynamic actor-oriented Markov models, the results show that in the European Parliament, male, senior parliamentarians from powerful member states receive more attention, and shared nationality is the strongest predictor of mentions. There is evidence of a reciprocity effect: activity in the debates increases other parliamentarians' participation in speeches.

Named-entity mentions in parliamentary speeches have also been used to obtain context windows for sentiment analysis. Person-name references in the transcripts of U.S. Congressional floor debates have been classified based on the context window around each mention to study sentiments towards proposed legislation~\cite{thomas2006get}. 
The mentions of political parties and the context around them in legislative speeches have been used to identify MPs who use polarizing rhetoric in the Norwegian Parliament~\cite{roed-2025-party-mentions}.

Networks of politicians in social media, especially on Twitter, are widely studied. Studies of different types of Twitter networks of politicians suggest that politicians are more likely to follow politicians with similar views and directly share their content, whereas they are more open to connecting with opposing ideologies by mentioning users~\cite{praet-2021-parl-twitter-networks}. For example, Twitter interactions by Members of the U.S. Congress were turned into networks and political affiliation, and ideology was studied~\cite{chamberlain-2021-us-twitter}. Directly sharing others' content leads to more polarised networks; that is, members of the same party are more likely to be in the same cluster than in networks based on mentions of other users.

There are also other forms of network analyses of parliamentary addresses. For example Bhattacharya analyzed networks of keywords to understand how party unity is achieved in plenary speeches in the German Bundestag~\cite{bhattacharya}. Ros and Wevers used networks of topics to portray the long-term evolution of debates in the Dutch parliament~\cite{ros2024epistemic}. J\"ackle and Metz studied the network of cosigning Oral Questions to the European Parliament~\cite{jackle2019oral}. (Oral Questions are formal tools of parliamentary scrutiny that allows Members of the European Parliament to hold other EU institutions---primarily the European Commission and the Council of the EU---directly accountable.)

\section{Methods}\label{sec_methods}

\subsection{Data collection and pre-processing}

The data contains speeches delivered in the Parliament of Finland by its members since its founding in 1907, as well as related metadata and links to external data sources~\cite{hyvonen_ps_data_2024}. The data is harmonised from previously available PDF, HTML, and XML files into unified, machine-readable formats. For the more recent speeches, starting from the yearly parliament session 2015, named entities such as person names, places, and organizations have been recognised and linked to corresponding resources~\cite{virkkunen2022finnish}. The data used in this study is available as an RDF knowledge graph, through a public SPARQL endpoint and as CSV or XML files~\cite{sinikallio_2025_15639893}.

We use the data from the electoral terms 2015--2018 and 2019--2022. During the electoral term 2015--2018 Finland had centre-right government, as Centre Party, National Coalition Party and the Finns Party formed a coalition government. In summer 2017, the Finns Party split into the Finns Party and the Finnish Reform Party. Finns Party moved to the opposition, and the Finnish Reform Party stayed in the government. For the later electoral term, centre-left government coalition consisting of the Social Democratic Party of Finland, Centre Party, Green League, Left Alliance, and Swedish People's Party of Finland was formed. In addition to the traditional parties mentioned above, three ``one man parties'' formed by individual MPs who left their original party during the electoral term were active: Movement Now, Seven Star Movement and Power Belongs to the People.

\begin{SCfigure*}[0.5]
    \centering
    \includegraphics[width=1.5\linewidth]{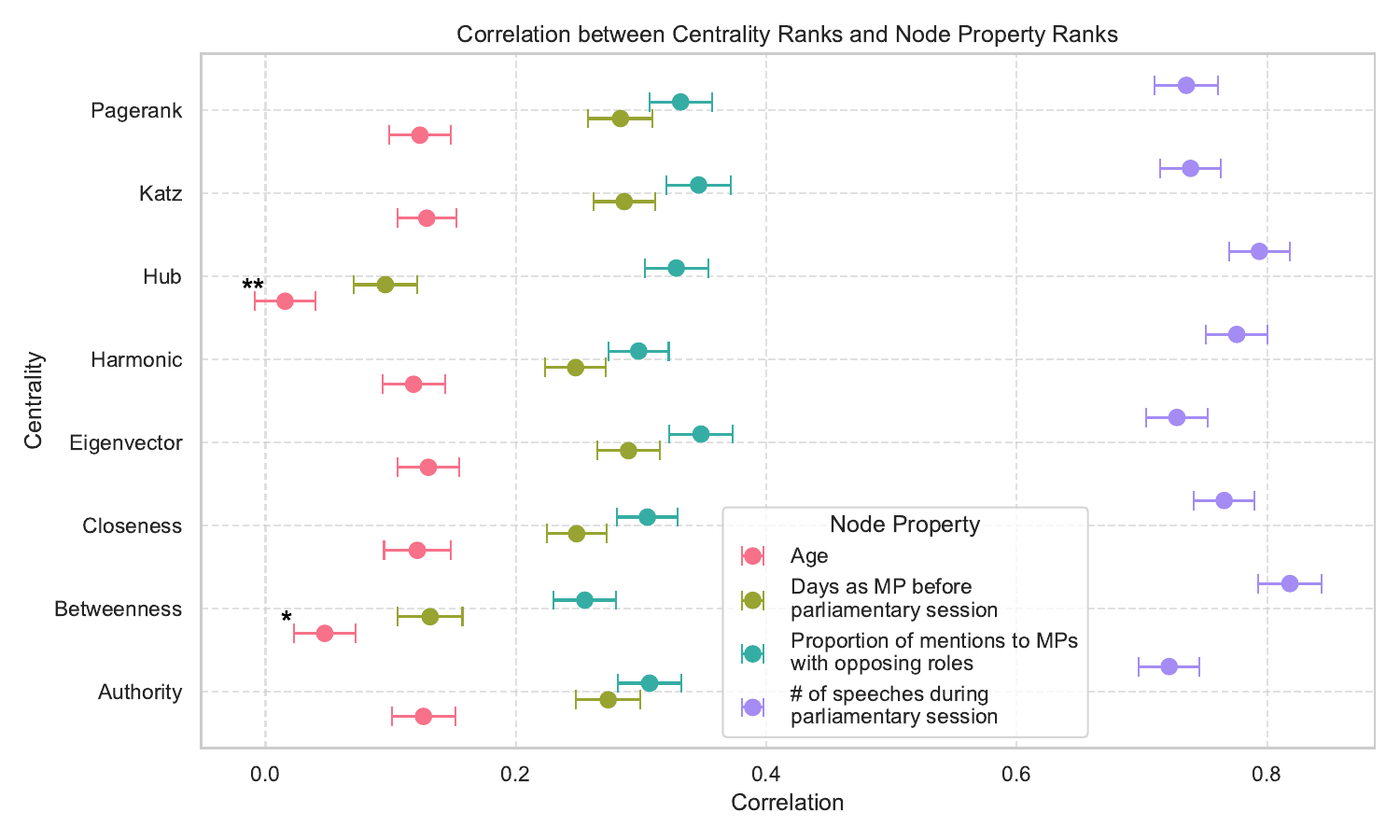}
    \caption{Correlation between centrality rankings and the rankings based on the age, political age, mentions across government/opposition lines and the number of speeches ($p<0.01$). *p-value $<0.03$ **p-value $\approx0.27$}
    \label{fig:centrality_attributes}
\end{SCfigure*}

\begin{figure}
    \centering
    \includegraphics[width=\linewidth]{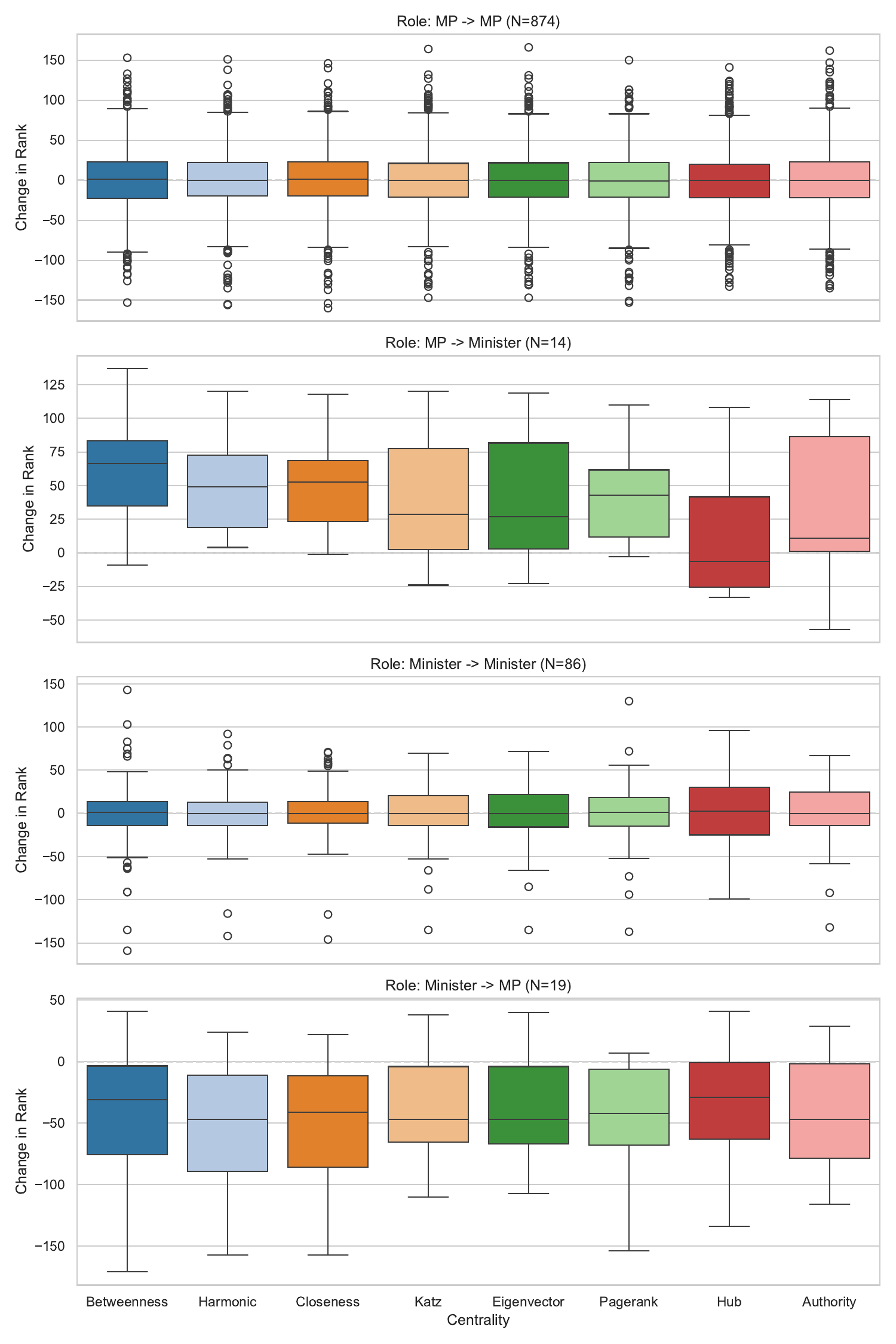}
    \caption{The impact of MPs staying as MP (top), becoming minister, staying as minister or losing minister position (bottom) on various measures of network centrality. The visualization contains standard box-and-whiskers plots showing the mean, first and third quartiles in the box and 1.5 times the interquartile range (the whiskers), and outliers.}
    \label{fig:ministers}
\end{figure}

We used SPARQL-queries based on matching graph patters~\cite{sparql_w3c} to obtain relevant data and construct a directed, weighted network of MPs for each yearly parliamentary session during electoral terms 2015--2018 and 2019--2022. Speeches delivered by the Speaker of the Parliament were not included, as they do not constitute actual debate. Edge weights correspond to the number of speeches in which the source node has mentioned the target node during the parliamentary session. 

The nodes are limited to people who were MPs during a parliamentary session for at least 30 days. Usually, MPs who have been active for less than 30 days have left the Parliament for one reason or another. For the MPs, additional metadata is queried. For example, name, number of speeches given during the parliamentary session, birthday, number of days as MP before the parliamentary session as ``political age'', political party of the MP at the end of the parliamentary session, proportion of total mentions directed to the MP's own party members, and government or opposition position. 

Networks consist of one weakly connected component and 0-3 isolates or MPs who have neither been mentioned nor mentioned anyone during the parliamentary session. These isolates are removed from the analysis. The resulting networks, considering all edges of non-zero weight, are relatively dense. To visualise such networks and to analyze them with binary network methods, it makes sense to prune the least important edges. Most straightforwardly, this would be done by thresholding---only including edges with weights above a certain threshold. However, different parts of the network may have varying levels of activity, meaning that identifying important edges (so-called \textit{backbone extraction}) would require an adaptively changing threshold. The most popular method for implementing that idea is the \textit{disparity filter}~\cite{serrano_backbone_2009}, which we use with the standard parameter value $\alpha=0.2$. After disparity filtering, some nodes end up isolated in the filtered network if none of their adjacent edges are recognised as significant, and are therefore not shown in visualizations. The visualizations were produced using Gephi \cite{bastian2009gephi}, and Yifan Hu~\cite{yifanhu2005} algorithm was used for the network layouts.

One limitation of our analyses is the quality of the named entity recognition and linking. Especially when the MPs share the same last name, the linking of the recognised name in question is sometimes mixed up. For example, during the electoral term 2015--2018 we had President Sauli Niinist\"o, Minister of Defence Jussi Niinist\"o, and Ville Niinist\"o who was the chair of the Green League party. Even if the names are recognised, if they are not linked properly, the number of mentions to MPs who share family names with other MPs can be overestimated. Also, only direct mentions of people's names are taken into account; MPs may sometimes refer to each other indirectly (e.g., ``the previous speaker'') rather than by name. 
While our results show that MPs do not segregate into structurally isolated network communities, this does not automatically imply an absence of rhetorical hostility. Future iterations of this work could integrate context-window sentiment analysis to distinguish between supportive cross-mentions (alliances) and adversarial call-outs (attacks).

\begin{SCfigure*}[0.5]
    \centering
    \includegraphics[width=1.5\linewidth]{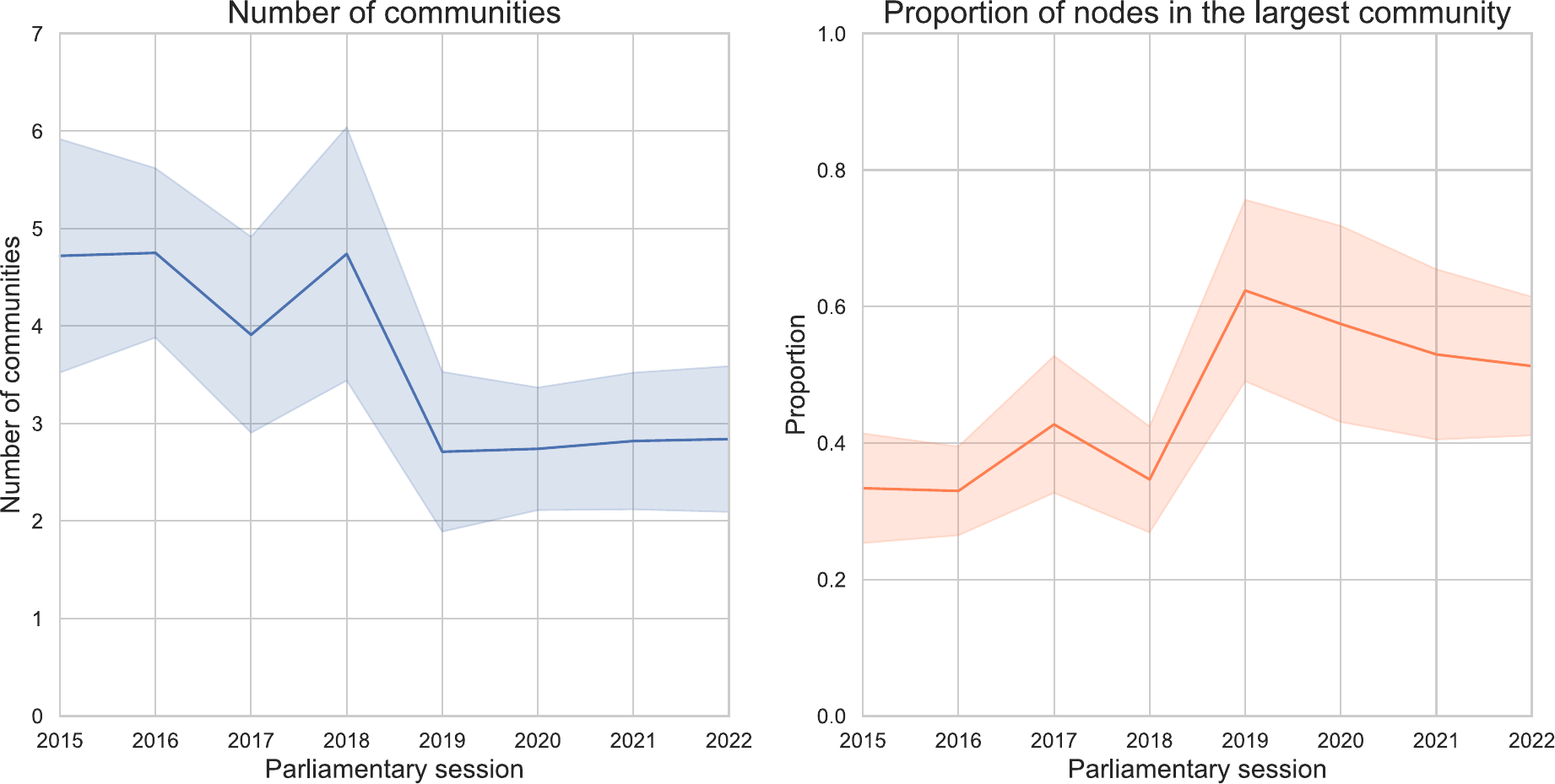}
    \caption{Mean number of communities obtained by the stochastic block model $\pm$ one SD and mean of the proportion of the nodes in the largest community $\pm$ one SD for $N=100$.}
    \label{fig:communities}
\end{SCfigure*}

\subsection{Centrality analysis}

To identify influential actors and map structural hierarchies within the Finnish Parliament, we computed centrality metrics for nodes for each yearly parliamentary session network. The importance of a node can be defined in different ways; centrality measures are the most common proxy for importance, defined by the network structure alone. Also, centrality measures come in many different flavors, each conceptualising an idea of what it means for a node to be central in the network~\cite{boldi-2014-harmonic-centr}. In real networks, there is usually a moderate or strong correlation between different centrality metrics~\cite{bi-2026-correlations-centrality}. Parliamentary name-mention networks characterise communication directionality (the speaker vs.\ the spoken-of) and frequency (the number of mentions). We utilise metrics that capture different structural dynamics.

We calculate closeness centrality~\cite{freeman1978closenesscentrality} and harmonic centrality~\cite{boldi-2014-harmonic-centr} to measure how efficiently an MP can reach or be reached by all other actors through shortest paths in the network. All nonzero edge weights are collapsed to 1.
We add betweenness centrality~\cite{freeman1977betweennesscentrality} for weighted networks to quantify the frequency with which an MP sits on the shortest paths connecting other pairs of legislators, indicating an actor’s ability to act as a mediator or structural bridge between potentially disconnected factions~\cite{himelboim2014betweenness}.

To track broader reputational status and diffuse authority, we calculate eigenvector centrality, which scales an MP’s significance proportionally to the collective prominence of their immediate neighbors~\cite{bonacich2007some}. To correct for localised structural anomalies in directed graphs, such as source-node components without incoming edges, we calculate PageRank~\cite{page1999pagerank} and Katz centrality~\cite{katz1953centrality}, which provide more stable global influence scores by incorporating damping factors. The edge weights are taken into account for these centralities.

Finally, we utilise the Hyperlink-Induced Topic Search (HITS) algorithm for unweighted versions of directed networks to determine mutually reinforcing dual structural profiles: hub and authority weights~\cite{kleinberg-1999-hits}. In a political debate ecosystem, an MP with high hub centrality represents an active speaker who explicitly mentions and targets multiple key parliamentary figures. Conversely, an MP with high authority centrality reflects a visible figure who is systematically mentioned, addressed, or contested by numerous other MPs, regardless of their own individual speaking activity.

To perform longitudinal comparisons across sequential yearly parliamentary sessions despite minor variations in the number of MPs, absolute centrality coefficients are converted into standardised ordinal ranks within each discrete network. This normalization allows us to track systematic rank fluctuations relative to structural properties, capturing shifts alongside variations in political age, parliamentary floor activity, and transitions between ministerial and regular MP roles.

\subsection{Cohesion}

Social networks tend to exhibit assortative mixing, that is, nodes tend to be more likely to be connected to other nodes with similar attributes~\cite{newman_mixing_2003}. (Note that, commonly, people talk about assortativity by degree, but here we use the general term.) While assortativity concerns the attribute similarity of nodes connected by an edge, for our analysis, it makes sense to relax this constraint and consider the similarity of nodes indirectly connected. By this logic, we expect the distance between members of the same party to be shorter than that of MPs of different parties~\cite{aiello2012friendship}, which we refer to as \textit{cohesion}. (By the \textit{distance} between two nodes, we mean the smallest number of edges among all paths connecting them, without taking into account the directions and weights of the edges.)

We define a \textit{cohesion coefficient} based on the shortest paths for this purpose:
\begin{equation} \label{eq1}
    \text{Cohesion} = \frac{D_{\text{different}}/n_{\text{different}} }{ D_{\text{same}} / n_{\text{same}}}
\end{equation}
Where $D_{\text{different}}$ is the sum of shortest distances between nodes who belong to different parties, $n_{\text{different}}$ is the number of such node pairs, $D_{\text{same}}$ is the sum of shortest distances between node pairs belonging to the same party and $n_{\text{same}}$ is the number of node pairs that are part of the same party. Cohesion is calculated for each parliamentary session, and by shuffling the party labels, we obtain p-values for them.

A cohesion value of one indicates that the political party of which a node is part does not have an effect on the distance between nodes. A cohesion value larger than one indicates assortativity with respect to political parties.

\subsection{Core-periphery structure}

As we will see later, network visualizations of weighted, directed mention networks suggest a possible core-periphery structure, with a single core surrounded by a periphery whose link density decreases with distance from the core. Core-periphery structure, originally formalised by Borgatti and Everett~\cite{borgatti2000models},  characterises a system partitioned into a central core block with highly interconnected nodes and a structurally sparse periphery block where nodes interact primarily through the core's intermediaries~\cite{borgatti2000models, zhang-2015-core-periphery}. 

We evaluate whether the network exhibits a core-periphery structure using stochastic block modeling (SBM), where nodes are divided into blocks that connect with each other with certain probabilities~\cite{peixoto-2019-sbm}. Some block models ignore variation in degree and end up dividing the network into blocks of higher and lower degree nodes, which is problematic in the case of real-world networks with broad degree distributions~\cite{karrer-2011-sbm-community}. However, for core-periphery detection, this behavior is preferable, as nodes in the core usually have a higher degree than nodes in the periphery~\cite{zhang-2015-core-periphery}.

If the largest community found by the stochastic block modeling includes a significant proportion of the nodes, there cannot be multiple cores (cf. Ref.~\cite{kojaku}). We also validate results against external attributes, specifically, political party affiliations and government versus opposition coalitions. We calculate precision, recall, and F1-scores for the communities and external attributes, where precision describes the homogeneity of the community, recall describes the compactness of it, and the F1-score is the harmonic mean of the two~\cite{cherepnalkoski2016retweet}. Multiple other measures for evaluating community detection are listed in~\cite{liu2019evaluation}. For each yearly session graph, we record the structural blocks discovered by the SBM, uniformly shuffle the external party and coalition labels across nodes, and recompute precision, recall, and F1 scores. By repeating these permutations, we calculate empirical p-values for precision, recall, and F1-score.

\subsection{Motif analysis}

The fact that our network is directed allows us to explore its motif structure. A \textit{network motif} is a small (typically two- or three-node) subgraph that is heavily overrepresented or suppressed compared to the expected number in a reference model~\cite{milo-2002-motifs} (as measured by a Z-score). These motifs are then (if overrepresented) typically interpreted as a functional component of the system. 

We count motifs in networks and Z-scores for them, using Pymfinder Python library~\cite{Mora-2018-pymfinder}, and focus on two types of motifs: dyads formed by two nodes and triads formed by three nodes. Only the directions of edges are considered, nonzero weights are collapsed to 1. Positive Z-scores imply over-representation of the motif in the network over a random graph with the same degree distribution; negative Z-scores imply under-representation~\cite{prill2005motifs}.

There are two types of dyads: one in which the link goes in one direction between two nodes, or one in which the edge goes in both directions between two nodes. In our case, bidirectional links indicate that MPs counter-mention or validate each other. We expect parliamentary name-mention networks to have higher reciprocity than in random null models. We also calculate overall reciprocity for the networks. For a directed network, there are 13 different types of triads~\cite{Mora-2018-pymfinder}. We study whether some of the triads are significant and whether there are differences across parliamentary sessions or electoral terms.

\section{Results}\label{sec_results}

\subsection{Statistics}

We start our analysis by visualizing the weighted, directed parliamentary-mention networks and their time evolution through parliamentary sessions; see Fig.~\ref{fig:networks}. One immediate impression (that we will corroborate later) is that the networks have one dense core and a sparse periphery. While nodes with the same color (political party) tend to be closer to one another, indicating assortativity, we also note that nodes are not polarised into near-disconnected subnetworks (as observed in studies of political blogs~\cite{political_blogs} and Twitter retweet networks~\cite {praet-2021-parl-twitter-networks}). 

Table~\ref{tab:parl_ses_statistics} shows basic information about the parliamentary sessions and the corresponding networks. The opening parliamentary sessions of the electoral terms (2015, 2019) are shorter due to elections, resulting in fewer speeches and mentions. Consequently, the density and overall reciprocity of the corresponding network are also lower. The overall reciprocity of the networks is moderately high. In the case of the European Parliament, there was evidence of a reciprocity effect, that is, the parliamentarians were more likely to mention people who had mentioned them before~\cite{walter_ep_mentions_2023}. Similar signs of debate can also be seen in the Finnish Parliament. Networks consist of one weakly connected component and a few isolates (MPs who have not mentioned other MPs or have not been mentioned during the parliamentary session). Isolates are not taken into account in the analyses. The isolated nodes after the disparity filter are not shown in Figure~\ref{fig:networks}.

\begin{SCtable*}[0.66667]
    \centering
    \begin{tabular}{c|r|c|c|c|c|c|c}
    \makecell{Parliamentary \\ session} & Speeches & Mentions & MPs & Density & \makecell{Overall \\ reciprocity} & Isolates & \makecell{Isolates \\ after filter} \\
    \hline
    2015 & 8669 & 3860 & 201 & 0.10 & 0.38 & 1 & 36 \\
    2016 & 14027 & 5295 & 202 & 0.13 & 0.46 & 0 & 17 \\
    2017 & 13637 & 5345 & 204 & 0.13 & 0.45 & 2 & 17 \\
    2018 & 15628 & 5983 & 206 & 0.14 & 0.43 & 1 & 13 \\
    \hdashline
    2019 & 6931 & 3841 & 206 & 0.09 & 0.37 & 2 & 32 \\
    2020 & 12483 & 5175 & 202 & 0.13 & 0.43 & 3 & 20 \\
    2021 & 12460 & 5153 & 205 & 0.12 & 0.43 & 1 & 31 \\
    2022 & 12832 & 4822 & 203 & 0.12 & 0.44 & 2 & 18 \\

    \end{tabular}
    \caption{Number of speeches (excluding speeches of the Speaker of the Parliament), name mentions and active MPs during each parliamentary session. For the corresponding networks, the density, overall reciprocity, and number of isolates before and after the disparity filter are listed.}
    \label{tab:parl_ses_statistics}
\end{SCtable*}

\subsection{Assortativity}

As seen in Fig.~\ref{fig:networks}, nodes of the same color (political party) tend to be close to one another, indicating assortativity among party members. Figure~\ref{fig:coefficients} shows the cohesion coefficients measuring assortativity for each parliamentary session, when the weight and direction of the links were not taken into account. For every parliamentary session except the 2016 session, the coefficient exceeds one, indicating significant assortativity. Coefficients are slightly higher for the electoral term 2019--2022 than for the previous electoral term, indicating that MPs mention members of their own party more often. 

Figure~\ref{fig:party_self_mentions} shows the proportion of all mentions made by party members towards the members of their own party, divided by the proportion of the party members out of all the MPs. A value of one indicates that the party mentions its own members as expected, assuming the mentions are uniformly distributed across all MPs. For example, five MPs of the Christian Democrats party elected for both electoral terms form a tightly-knit group who mention each other often, probably supporting the party's agenda. After the split of the Finns Party in 2017, the remaining members of the Finns Party rely on each other in the form of mentions. 

The members of parties in the 2018--2022 left-centre coalition government (Left Alliance, Green League, Social Democrats, and Swedish People's Party of Finland) tend to mention fellow party members more when they move from opposition to the government. The number of self-mentions by the Centre Party increases as it loses the title of the largest party and the prime minister position between electoral terms. The opposite happens with the National Coalition Party and Finns Party, which are considered to be on the right side of the political spectrum; they tend to have more self-mentions when in Opposition. This could explain the differences in cohesion between electoral terms apparent in figure~\ref{fig:party_self_mentions}.

\subsection{Centrality analysis}

To study how different attributes affect MPs' centrality, we calculated several centrality measures for the MPs, ranked them, and compared the results with age, political age, the proportion of mentions across opposition and government, the number of speeches, and ministerial status. Correlations between centrality rankings and the rankings of some node attributes are shown in Fig.~\ref{fig:centrality_attributes}. For random null models, node attributes were shuffled. 

People who have given more speeches are naturally more central, as they have had more opportunities to mention others, are more likely to be mentioned in others' answers, and have overall better visibility. Political seniority has a moderate positive correlation with centrality rankings, and it matters more than the age of the MP. Results from the political-mention networks in the European Parliament also indicate that the attention is on senior members \cite{walter_ep_mentions_2023}. The proportion of mentions to MPs with opposite roles, whether in government or opposition, has a moderate positive correlation with centrality metrics. This indicates that MPs gain visibility by mentioning other MPs across party and government coalition lines, prompting potential debates. 

According to Fig.~\ref{fig:ministers}, the centrality rankings tend to rise between parliamentary sessions when an MP attains minister status, and fall when a minister becomes an ordinary MP. When the status stays the same, the mean rank change is zero, but for some individuals, the change can be very large. Questions related to the cases each ministry is handling are probably often directed to the minister in charge, increasing the number of mentions directed towards them.

\subsection{Core-periphery structure}

The core-periphery detection algorithms did not indicate a significant core-periphery structure with a single core, as the visualizations suggest. Instead, we tried community detection with stochastic block modeling. The left side of Fig.~\ref{fig:communities} shows the mean number of communities in different parliamentary sessions, while the right side of Fig.~\ref{fig:communities} shows the mean proportion of nodes in the largest community during parliamentary sessions. The number of communities tends to be lower, and the size of the largest community tends to be larger, during the later electoral term, indicating a clearer core-periphery structure.

Comparisons of the communities to party and government role labels gave poor precision and recall (see tables~\ref{tab:community_precision} and~\ref{tab:community_recall}), which were often not significantly higher than the mean precision and recall of the networks where labels were randomised. Especially the recall and F1-scores are not often significant. Although some homogeneity is detected, especially from the later electoral term, and party members have some bias towards mentioning their own party members, that bias does not lead to clearly separated communities. 

In the sense of person-name mentions, the discussions in the Parliament of Finland are not heavily polarised. A closer look at mentions between the government and opposition shows that 45-53\% of mentions by government MPs across parliamentary sessions refer to opposition MPs. On the other hand, 59-73\% of mentions by opposition members target government MPs, emphasizing the opposition's role to challenge the government.

\begin{table}[]
    \centering
\begin{tabular}{l|lll|}

Year & Precision & Null-model precision  & p-value \\
\hline
 2015 & 0.635 & 0.536 & 0.001 \\
 2016 & 0.536 & 0.533 & 0.278 \\
 2017 & 0.510 & 0.510 & 0.389 \\
 2018 & 0.514 & 0.508 & 0.130 \\
 2019 & 0.602 & 0.521 & 0.001 \\
 2020 & 0.547 & 0.519 & 0.002 \\
 2021 & 0.574 & 0.517 & 0.001 \\
 2022 & 0.526 & 0.519 & 0.124 \\

\end{tabular}
    \caption{Comparison of communities to government/opposition roles. Parliamentary session, the precision for the original graph, mean precision for the graphs with randomised labels ($N=100$), and p-values.}
    \label{tab:community_precision}
\end{table}

\begin{table}[]
    \centering
\begin{tabular}{l|lll|}

Year & Recall & Recall mean & p-value \\
\hline
 2015 & 0.484 & 0.478 & 0.679 \\
 2016 & 0.225 & 0.224 & 0.509 \\
 2017 & 0.399 & 0.378 & 0.482 \\
 2018 & 0.401 & 0.399 & 0.659 \\
 2019 & 0.504 & 0.465 & 0.633 \\
 2020 & 0.582 & 0.589 & 0.714 \\
 2021 & 0.707 & 0.678 & 0.801 \\
 2022 & 0.500 & 0.504 & 0.548 \\

\end{tabular}
    \caption{Comparison of communities to government/opposition roles. Parliamentary session, the recall for the original graph, mean recall for the graphs with randomised labels ($N=100$), and p-values.}
    \label{tab:community_recall}
\end{table}

\begin{SCfigure*}[0.5]
    \centering
    \includegraphics[width=1.5\linewidth]{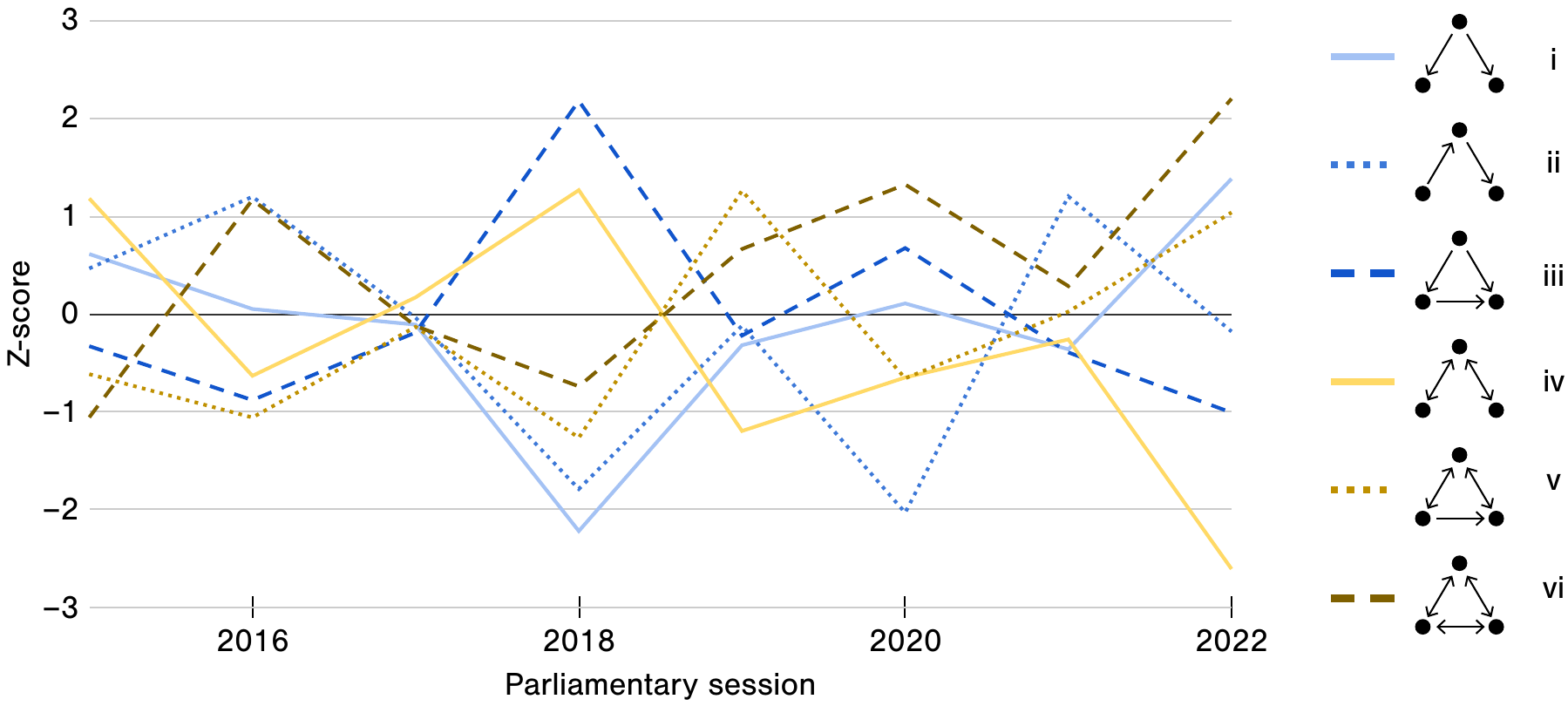}
    \caption{Z-scores showing the over- and under-representation of six (of thirteen possible) directed, connected three-vertex subgraphs. The motifs (i--vi) (identified using pymfinder \cite{Mora-2018-pymfinder}) shown on the left are diagrammed on the right.}
    \label{fig:zscores}
\end{SCfigure*}

\subsection{Motif analysis}

Turning to our motif analysis, the Z-scores for dyads indicate strong and significant reciprocity compared to a random null model. For the electoral term 2015--2018, Z-scores for bidirectional edges vary between 25.3 and 34.7, and for the electoral term 2019--2022 between 22.2 and 28.2, mirroring the overall reciprocity shown in table~\ref{tab:parl_ses_statistics}. The first parliamentary session of the electoral term is shorter due to elections, which means fewer speeches and therefore lower density and overall reciprocity of the related network, as well as smaller Z-scores for the bidirectional edges. The results are in line with other real-world social networks where the reciprocity tends to be larger than expected compared to random null models~\cite{jiang-2015-reciprocity}.

The Z-scores indicated some significant triads, but those triads were not significant throughout the parliamentary sessions, as seen in figure~\ref{fig:zscores}. The larger Z-scores usually appear in the last parliamentary session of the electoral term. In the latter electoral term, the complete triad (vi) and triad v often have positive Z-scores, resulting in a lower appearance of triad iv. Social networks usually express higher clustering coefficients measuring how tightly the neighbors of a node are connected~\cite{mislove-2007-online-social}, and several clustering metrics have also been defined for directed networks \cite{trolliet-2021-directed-clustering}.

\section{Discussion}\label{sec_disc}

 The structural analysis of the parliamentary name-mention networks in the Parliament of Finland provides new computational insights into the behavioral patterns of the parliamentary discussion. By mapping interactions over two distinct and contrasting electoral terms (2015–2018 and 2019–2022), we study how changing government coalitions and the roles of individual MPs shape political discourse.

 There is a high level of systemic reciprocity across all parliamentary sessions, backed by the overall reciprocity and highly significant dyadic Z-scores. This suggests that parliamentary mention networks operate under a ``counter-mention'' or conversational validation paradigm. This is consistent with findings from the European Parliament, where mentioning the name later increases the presence of parliamentarians in other representatives' speeches~\cite{walter_ep_mentions_2023}, but in stark contrast to the observations of polarization on social media~\cite{praet-2021-parl-twitter-networks,political_blogs}, and to some extent to parliamentary interaction in the UK~\cite{peterson-2018-polarization} and Norway~\cite{cherepnalkoski2016retweet}.

 While social media networks of politicians---such as Twitter retweets~\cite{praet-2021-parl-twitter-networks} or political blogospheres~\cite{political_blogs}---showcase polarised echo chambers and highly disconnected components, the Finnish parliamentary floor presents a more unified topology. Despite a slight bias towards mentioning members of one's own party, instead of fracturing into polarised camps, the Finnish Parliament maintains a connected ecosystem. The government has to maintain unity, while the opposition challenges it, leading to name-calling between allies and opponents. Opposition MPs direct the vast majority of their mentions toward the ruling government, explicitly reinforcing their democratic mandate to challenge executive policies and offer viable alternatives (cf.\ Ref.~\cite{sagarzazu2017coalition}). Conversely, government MPs dedicate roughly half of their naming focus to opposition actors, illustrating a balanced dynamic of defense, counter-argument, and direct cross-coalition debate. Furthermore, transitions between government and opposition status fundamentally alter how parties mention their own members.

The centrality analysis highlights the structural mechanisms that drive political visibility and authority on the floor. While the raw volume of speeches unsurprisingly remains the strongest predictor of network centrality, political seniority exhibits a moderate, stable positive correlation (echoing Ref.~\cite{walter_ep_mentions_2023}), and mentions across the government/opposition divide tend to increase the MP's centrality. Attaining minister status triggers a significant spike across nearly all centrality metrics, particularly in authority weight (supporting the United Kingdom and Norway observations of Ref.~\cite{yildirim-2023-parl-activities-media}). Ministers serve as the primary targets for policy questions and scrutiny during Question Time, drawing heavy incoming edges from both the opposition and their own coalition peers. When these individuals return to being regular MPs, their structural authority promptly collapses back to the baseline.

We hope our study will inspire further investigation in political mention networks. Two obvious directions to go are: toward comparative analyses, extending beyond Finnish politics, and to include more diverse data to try to achieve a clearer and more nuanced picture.

\bibliographystyle{abbrv}
\bibliography{sn-bibliography}

\end{document}